\documentclass[a4paper,fleqn,usenatbib]{mnras}
\usepackage[T1]{fontenc}
\usepackage{ae,aecompl}
\usepackage{graphicx}   % Including figure files
\usepackage{amsmath}    % Advanced maths commands
\usepackage{amssymb}    % Extra maths symbols

\title[The LITE study in SMC eclipsing binaries]{The first study of the Light-Travel Time Effect in bright eclipsing binaries in the Small Magellanic Cloud\thanks{Based
  on data collected with the Danish 1.54-m telescope at the ESO La Silla Observatory.}}

\author[P. Zasche]{
P. Zasche,$^{1}$\thanks{E-mail: zasche@sirrah.troja.mff.cuni.cz} M. Wolf,$^{1}$ J.
Vra\v{s}til$^{1}$\\
$^{1}$ Astronomical Institute, Charles University, Faculty of Mathematics and Physics, CZ-180~00
Praha 8, V~Hole\v{s}ovi\v{c}k\'ach 2, Czech Republic}

\date{Accepted XXX. Received YYY; in original form ZZZ}

\pubyear{2016}

\begin{document}
\label{firstpage}
\pagerange{\pageref{firstpage}--\pageref{lastpage}}
\maketitle

% Abstract of the paper                                18 radku
\begin{abstract}
The first 100 brightest eclipsing systems from the Small Magellanic Cloud (SMC) were studied for
their period changes. The photometric data from the surveys OGLE-II, OGLE-III, OGLE-IV, and MACHO
were combined with our new CCD observations obtained using the Danish 1.54-meter telescope (La
Silla, Chile). Besides the period changes also the light curves were analysed using the program
{\sc PHOEBE}, which provided the physical parameters of both eclipsing components. For fourteen of
these systems the additional bodies were found, having the orbital periods from 2 to 20~ years and
the eccentricities were found to be up to 0.9. Among the sample of studied 100 brightest systems we
discussed the number of systems with particular period changes. About 10\% of these stars show
eccentric orbit, about the same number have third bodies and about the same show the asymmetric
light curves. %All of these results came solely from the photometric observations of the systems and
%can be considered as a good starting point for the future dedicated observations.
\end{abstract}


%%%%%%%%%%%%%%%%%%%%%%%%%%%%%%%%%%%%%%%%%%%%%%%%%%

%%%%%%%%%%%%%%%%% BODY OF PAPER %%%%%%%%%%%%%%%%%%

\section{Introduction}

The classical eclipsing binaries still play a crucial role in modern astrophysics. We can study the
eclipsing binary (hereafter EB) light curve, and model its shape, revealing the physical parameters
of both eclipsing components as well as their mutual orbit (see e.g.
\citealt{2009ebs..book.....K}). It is still the most precise method to derive the individual
masses, radii, and luminosities of components (see e.g. \citealt{2012ocpd.conf...51S}).

The same apply also for the role of the EBs outside of our Milky Way Galaxy, however obtaining good
observations is much more tricky and time-consuming due to their low brightness. Hence, we usually
deal with a lack of data for analysis. This aspect has changed rapidly during the last two decades
thanks to the large photometric surveys like OGLE and MACHO. Owing to the long-lasting photometric
monitoring of the Magellanic Clouds (OGLE II, III and IV cover 16 seasons), almost 50000 EBs are
known outside our Galaxy \citep{2016AcA....66..421P}, and many of them are interesting enough for
further more detailed analysis. Hence, here comes our contribution to the topic.

As a well-known fact, the Magellanic Clouds have slightly different metallicity than our Milky Way
Galaxy (see e.g. \citealt{1997macl.book.....W}, or \citealt{2015ApJ...806...21D}). Therefore, we
can study whether this effect plays a role in eclipsing binary research, whether it is traceable in
the models, or whether our data are sufficiently precise to distinguish between models with
different metallicities. We can also study the stellar multiplicity in general -- Is the frequency
of multiple systems the same in Magellanic Clouds as in our own Galaxy? \cite{2016MNRAS.455.4136B}
show quite recently that of about 1/12 of all eclipsing binaries observed by the Kepler space
telescope probably contain additional components that can be detected only via eclipse timing
variations. Is this number roughly the same in other stellar sample, even outside Milky Way?

\section{The system selection}

Continuing our similar study of period changes in eight eclipsing binaries located in LMC
\citep{2016A&A...590A..85Z}, we now moved our attention to the SMC galaxy and another approach. Due
to the well-known fact that the probability of detecting another third component in eclipsing
binary strongly depends on the primary mass (see e.g. \citealt{2013ARA&A..51..269D}), we focused on
the most massive stars (hence also the most luminous ones from the OGLE survey). Therefore, we took
the first one hundred brightest targets in the OGLE III survey from the SMC galaxy (more precisely:
in the $I$ filter) and performed the analysis of its period changes. Additionally, some targets
were also added to this sample as by-chance discoveries in some of our monitored fields.

The selection criterion based on the data quality was as follows. We have chosen only such systems,
which have the depths of their minima deeper than the scatter of the light curve itself

Another selection criterion was the data coverage. Because we focused on periodic variations in the
$O-C$ diagrams, we decided to include in our study those systems, which have at least one period of
the variation already covered (either with the survey data or our own observations).

\section{The analysis}

The method was the same as in our previous paper \citep{2016A&A...590A..85Z}, which means that the
whole time interval was divided into several seasons and the light curve template from the {\sc
PHOEBE} (see below) fit was used for deriving the individual times of minima. If there was some
obvious change of its orbital period, then the system was classified as a suspicious and analysed
in more detail. This means that also the other available photometry was collected, mostly the {\sc
Macho} \citep{2007AJ....134.1963F}, {\sc OGLE II} \citep{2004AcA....54....1W}, and {\sc OGLE III}
\citep{2013AcA....63..323P}, and {\sc OGLE IV} \citep{2016AcA....66..421P} databases. Moreover, for
some of the systems we also collected some new data using the Danish 1.54-meter telescope located
at the La Silla Observatory in Chile. The data mining from all these data sources together with our
new photometry led to the selection of fourteen interesting systems showing periodic modulation of
their orbital periods.

The light curve (hereafter LC) analysis was carried out using the program {\sc PHOEBE}
\citep{2005ApJ...628..426P}, based on the Wilson-Devinney algorithm \citep{1971ApJ...166..605W} and
its later modifications. We used the OGLE III data for the light curve modelling, because these are
typically of the best quality, obtained over the longer time span and the phase light curves are
well-covered.

The {\sc PHOEBE} code enables us to construct the theoretical LC, which is later used as a template
to derive the times of eclipses. Hence, our LC fit needs to be as precise as possible. For all of
our fourteen system we found that their orbits are circular, hence the eccentricity was fixed at
zero. For the starting ephemerides, we used the same ones as published by
\cite{2013AcA....63..323P} in their catalogue and later modified according to the period changes.
The primary temperatures were derived from the published photometric indices by
\cite{2002ApJS..141...81M}, and \cite{2002AJ....123..855Z}. For only a few systems their spectral
types or primary temperatures were published, then we used these values, of course. See Table
\ref{InfoSystems} for more information about the individual systems in our sample.

Therefore, the set of the fitted quantities was the following: The temperature of the secondary
component $T_2$, the inclination angle $i$, the Kopal's modified potentials $\Omega_i$, and the
luminosities of the components $L_i$. Having no information about the radial velocities, the mass
ratio can only hardly be derived (see e.g. \citealt{2005Ap&SS.296..221T}), hence we fixed it to
$q=1.0$. The limb darkening coefficients were interpolated from the van Hamme's tables
\citep{1993AJ....106.2096V}, and the synchronicity parameters ($F_i$) were also kept fixed at
values of $F_i = 1$. Because we deal with very hot stars here, we also fixed the albedo
coefficients $A_i$ at a value 1.0, as well as the gravity darkening coefficients $g_i = 1.0$.

For studying the apparent variations of the orbital period in these binaries, we used a well-known
"Light-Travel Time Effect" (or LTTE) hypothesis \citep{Irwin1959}. It is based on the assumption
that the two eclipsing stars are being accompanied by some hidden distant component, orbiting
around the common barycenter with the eclipsing pair. Hence, we deal with a classical hierarchical
system. As the pair moves around a common center of mass, the eclipses of the binary occur earlier
or later depending on the current position of the stars with respect to the observer. For some
discussion and limitations of this method see for instance \cite{Mayer1990}. A similar method was
used quite recently for discovering several dozens of new triple systems in the Kepler field, see
\cite{2016MNRAS.455.4136B} or \cite{2015AJ....150..178G}. As far as we know, there was no other
study analysing the period changes via LTTE located in the SMC published until yet, only the paper
studying the 90 systems with apsidal motion by \cite{2016MNRAS.460..650H}.

The times of minima for the analysed systems were derived using the AFP method presented in
\cite{2014A&A...572A..71Z}. This method uses the LC template as derived from the {\sc PHOEBE} and
shifts the template in both $x$ and $y$ axes together with the phased light curve in the particular
dataset to achieve the best fit. These datasets were constructed according to the quality and
density of the data (for the OGLE this usually means one dataset per one year of observations, but
for short-periodic variables more fine division into datasets was used). Using the MACHO, OGLE and
our new data we obtained minima times spanning over many years. Hence, detecting the orbital
periods of the order of a few years to a decade was possible.

The whole fitting process was performed in several steps. At first the ephemerides from
\cite{2013AcA....63..323P} were used and the preliminary solution was found in {\sc PHOEBE} (using
typically one season of OGLE data with the best coverage). With this light curve template the AFP
method produced some preliminary minima times and we were able to see whether the system was
suitable for a further analysis or not. The second step was the $O-C$ analysis, refining the
orbital period which was then used in {\sc PHOEBE} for a more detailed modelling of the light
curve. With the final light curve template the final times of minima were derived and the $O-C$
analysis was performed. Different kinds of modulations of the orbital period can be studied in this
way, but we focused only on the periodic modulation. Sometimes our approach only led to improvement
of the linear ephemerides and no variation was found. Sometimes, some other phenomena were found,
but such systems are not of our interest and these systems are only briefly summarized below in
Section \ref{stat}.

However, at this place we have to emphasize that the solutions presented are only mathematical
ones, and especially the errors (e.g. from {\sc PHOEBE}) can sometimes be rather underestimated.
More discussion about the conclusiveness of the fits are given below in Section \ref{discussion}.

\begin{figure*}
  \centering
  \includegraphics[width=0.7\textwidth]{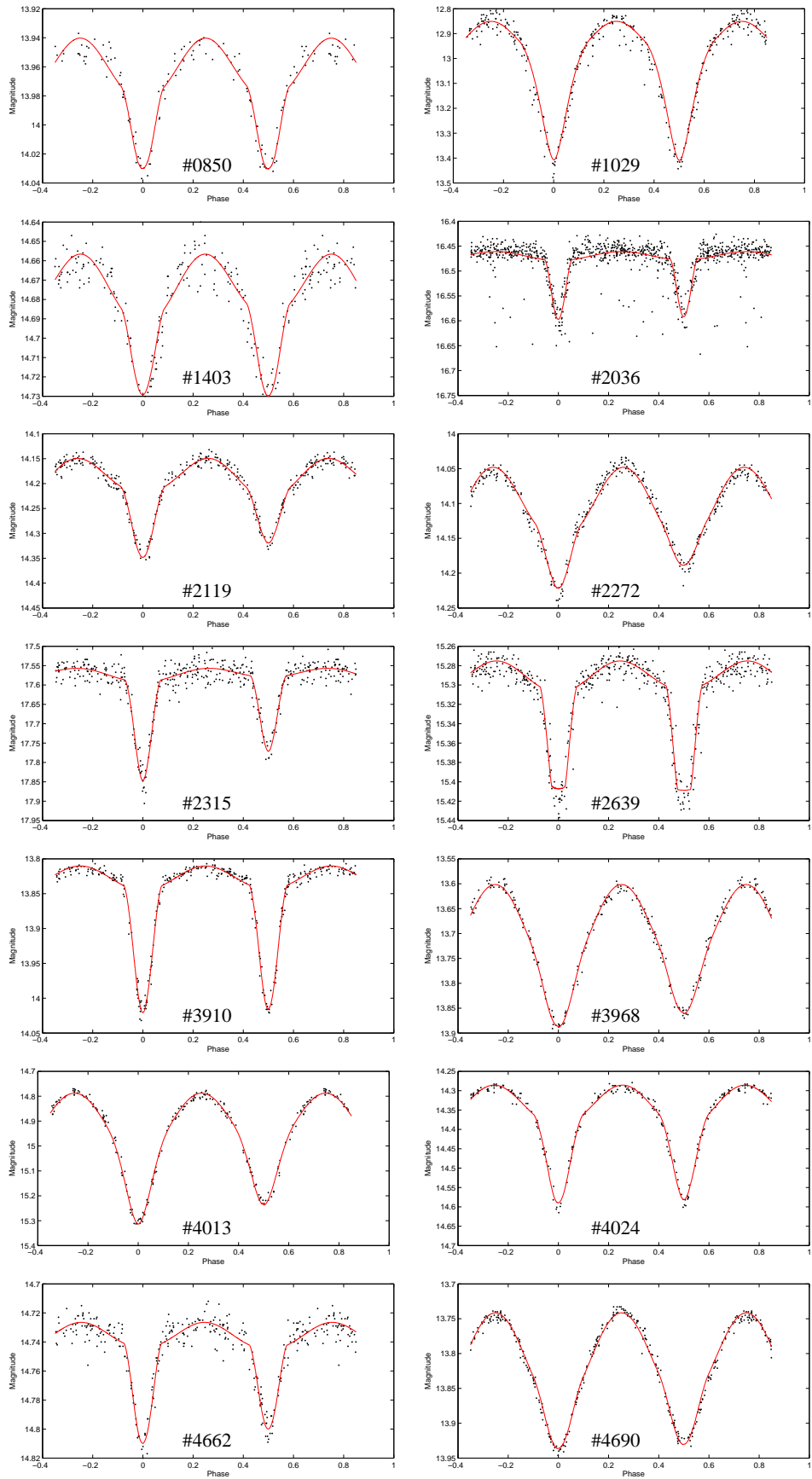}
  \caption{Plot of the light curves of the analysed systems.}
  \label{FigLC}
\end{figure*}

\section{Individual systems}

For all of the tables and pictures below we decided to use an abbreviation of the long OGLE III
names. Therefore, instead of, for instance, OGLE-SMC-ECL-0850, we use only the $\#$0850 for a
better clarity.

Because the method of analysis is the same for all of the studied systems and the results are
sometimes similar, we focus only on the most interesting ones in our sample and discuss them in
more detail in the following subsections. The results are summarized in Tables \ref{InfoSystems},
\ref{LCparam}, and \ref{OCparam}, while the final plots are given in Figures \ref{FigLC}, and
\ref{FigOC}.

\subsection{OGLE-SMC-ECL-1403}

The system OGLE-SMC-ECL-1403 was found to be the detached one, with both rather hot components.
Quite surprising was high mass function of the predicted third body $f(m_3)$, which resulted from
rather short period $p_3$. However, even such solution is plausible, because also very high level
of the third light in the LC solution $L_3$ was found.

\subsection{OGLE-SMC-ECL-2036}

A similar situation is also for the system OGLE-SMC-ECL-2036, where a high value of the mass
function $f(m_3)$ is substantiated by the very high level of the third light from the LC solution.
The obvious high scatter of the LC is caused by difficult reduction due to the close bright stars.

\subsection{OGLE-SMC-ECL-2119}

The system OGLE-SMC-ECL-2119 is one of a few already analysed and published systems.
\cite{2005MNRAS.357..304H} collected besides the photometry also 17 spectra, and the analysis
yielded that both eclipsing components are of O9 spectral type, with the mass ratio of 1.086. This
value was also kept fixed during our LC fitting, together with the fixed $T_1$ temperature. There
resulted that the third light is significant, but not dominant (the previous study was not taking
the $L_3$ contribution as a free parameter). Our solution of the period changes resulted in a
consistent figure with a significant third body on a 5-yr orbit.

%%   - nesmyslne reseni, trochu to pripomina 08823 z LMC
%%    - porovnat nase a publikovane reseni

\subsection{OGLE-SMC-ECL-2272}

This was the only system in our sample, which shows a small asymmetry of its LC, where primary and
secondary minima appear slightly shifted from their positions in 0.0 and 0.5 in phase. In detached
configuration this usually means that the system is eccentric, however here we deal with a
semidetached configuration, hence this explanation is odd. As was shown elsewhere (e.g.
\citealt{2011IBVS.5991....1Z}) the asymmetric light curves sometimes mimic the false eccentricity
also in contact binaries. Hence, we shifted both primary and secondary minima to one common
ephemerides and performed the period analysis. This approach is justified due to the fact that both
primary and secondary minima behave in the same way and the periodic modulation with the period of
about 7.6 yr is clearly seen in both of them.

\subsection{OGLE-SMC-ECL-3910}

Another case where at least some information can be found in the already published papers, see
\cite{2012ApJ...748...96M}. The authors provided information about the spectral types of the
components to be of O5+O7 with a very important remark that there are some triple lines visible in
the spectra, which are also shifting. Hence, our analysis of the LC together with the period
changes clearly confirms their finding.

\subsection{OGLE-SMC-ECL-3968}

The system with the shortest detected period of the LTTE of about 2 years, and also with the second
shortest period of the inner eclipsing pair.

\subsection{OGLE-SMC-ECL-4013}
In this case there was quite problematic calibration of the temperature $T_1$ due to the
ill-defined value of the photometric index $(B-V)_0$. Therefore, we simply fixed the $T_1$ to
10000~K.

\subsection{OGLE-SMC-ECL-4024}

The system OGLE-SMC-ECL-4024 was found to exhibit double periodic modulation in its $O-C$ diagram.
Hence, we used the LTTE hypothesis two times for a detailed description of the minima times
observations. The results are plotted in Fig. \ref{FigOC}, where both modulations are clearly
visible. %(figs (a) with a total fit, and (b) with only the shorter modulation after subtracting the 19-yr variation).
Pure LTTE combined with the quadratic ephemerides are not able to describe the data in detail.
Parameters of both LTTE variations are given on separate lines in Table \ref{OCparam}. Such systems
were seldom discovered in our Galaxy, but this is the first time any such system is being detected
outside of Galaxy.

\subsection{OGLE-SMC-ECL-4690}

\begin{figure*}
  \centering
  \includegraphics[width=0.8\textwidth]{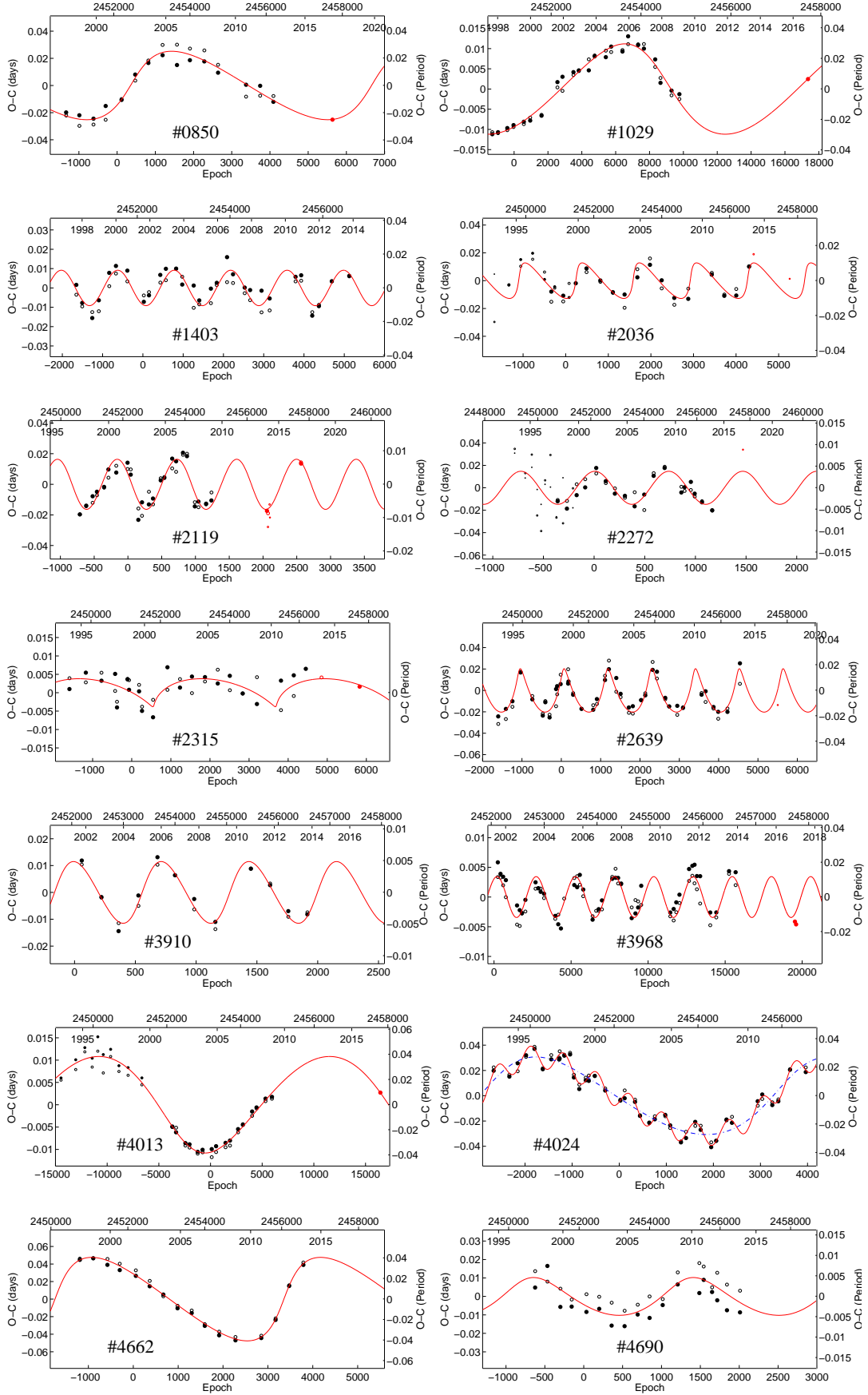}
  \caption{Plot of the $O-C$ diagrams of the analysed systems. The solid lines represent the final
  LTTE fit, filled points stand for the primary minima, while the open circles for the secondary ones,
  bigger the symbol, better the precision. The red data points represent our new observations.}
  \label{FigOC}
\end{figure*}

The last system in our set of stars also seems to be slightly asymmetric, a similar situation as
for $\#$2272. But here we ignored this asymmetry, which yielded to the $O-C$ diagram in Fig.
\ref{FigOC}, where the primary and secondary minima are slightly shifted, but both follow similar
period variation.

\begin{table*}
\caption{Identification of the analysed systems.}  \label{InfoSystems}
% \footnotesize
  \scriptsize
\begin{tabular}{lcccccccccccl}
   \hline\hline\noalign{\smallskip}
 System   & OGLE II$^{\,1}$ &  MACHO     &          RA           &             DE                              & $I_{\rm max}^{\,2}$ & $(V-I)^{\,3}$ & $(B-V)_0^{\,4}$  \\      %Teff        TeffFINAL
  \hline\noalign{\smallskip}

 OGLE-SMC-ECL-0850 & SC3 202715  &             & 00$^h$45$^m$18$^s$.20 & -73$^\circ$15$^\prime$23$^{\prime\prime}\!$.1 & 13.936 & -0.167 & -0.314 \\
 OGLE-SMC-ECL-1029 & SC4 88435   &             & 00$^h$46$^m$19$^s$.67 & -72$^\circ$50$^\prime$56$^{\prime\prime}\!$.7 & 12.858 &  0.796 & -0.009 \\
 OGLE-SMC-ECL-1403 & SC4 175130  &             & 00$^h$48$^m$17$^s$.96 & -73$^\circ$07$^\prime$19$^{\prime\prime}\!$.0 & 14.658 & -0.124 & -0.282 \\
 OGLE-SMC-ECL-2036 & SC5 306002  &208.16026.98 & 00$^h$51$^m$04$^s$.28 & -72$^\circ$47$^\prime$38$^{\prime\prime}\!$.9 & 16.463 & -0.075 & -0.144 \\
 OGLE-SMC-ECL-2119 & SC5 305884  &             & 00$^h$51$^m$20$^s$.18 & -72$^\circ$49$^\prime$43$^{\prime\prime}\!$.4 & 14.150 & -0.147 & -0.268 \\
 OGLE-SMC-ECL-2272 & SC6 77224   &208.16085.16 & 00$^h$51$^m$50$^s$.13 & -72$^\circ$39$^\prime$23$^{\prime\prime}\!$.1 & 14.048 & -0.151 & -0.285 \\
 OGLE-SMC-ECL-2315 & SC6 67920   &208.16084.320& 00$^h$52$^m$01$^s$.16 & -72$^\circ$44$^\prime$41$^{\prime\prime}\!$.8 & 17.572 & -0.073 & -0.105 \\
 OGLE-SMC-ECL-2639 & SC6 232226  &207.16140.25 & 00$^h$53$^m$07$^s$.01 & -72$^\circ$46$^\prime$21$^{\prime\prime}\!$.0 & 15.289 & -0.159 & -0.290 \\
 OGLE-SMC-ECL-3910 &             &             & 00$^h$59$^m$00$^s$.04 & -72$^\circ$10$^\prime$38$^{\prime\prime}\!$.1 & 13.825 &        & -0.322 \\
 OGLE-SMC-ECL-3968 &             &             & 00$^h$59$^m$20$^s$.47 & -71$^\circ$21$^\prime$42$^{\prime\prime}\!$.4 & 13.616 &        & -0.018 \\
 OGLE-SMC-ECL-4013 &             & 211.16529.5 & 00$^h$59$^m$31$^s$.20 & -73$^\circ$26$^\prime$56$^{\prime\prime}\!$.0 & 14.782 &        & -0.358 \\
 OGLE-SMC-ECL-4024 & SC8 129157  & 211.16539.2 & 00$^h$59$^m$34$^s$.19 & -72$^\circ$46$^\prime$57$^{\prime\prime}\!$.9 & 14.292 & -0.218 & -0.278 \\
 OGLE-SMC-ECL-4662 & SC9 163573  &             & 01$^h$03$^m$13$^s$.97 & -72$^\circ$25$^\prime$07$^{\prime\prime}\!$.6 & 14.737 & -0.214 & -0.323 \\
 OGLE-SMC-ECL-4690 & SC9 175323  &             & 01$^h$03$^m$21$^s$.30 & -72$^\circ$05$^\prime$38$^{\prime\prime}\!$.2 & 13.740 & -0.183 & -0.313 \\ \hline
\end{tabular}
\\
\begin{minipage}{0.9\textwidth}
% \scriptsize
Notes: [1] - The full name from the OGLE II survey should be OGLE LMC-SCn nnnnnn, [2] - Value taken
from \cite{2013AcA....63..323P}, [3] - Value taken from \cite{1998AcA....48..147U}, [4] Dereddened
value derived from the (B-V) and (U-B) indices taken from \cite{2002ApJS..141...81M} or
\cite{2002AJ....123..855Z}.
\end{minipage}
\end{table*}

\begin{table*}
\caption{Light curve parameters for the analysed systems, the results from {\sc PHOEBE}.}
\label{LCparam}
 \scriptsize
\begin{tabular}{lccccccccccc}
  \hline\hline\noalign{\smallskip}
  System  & $T_1$ (fixed) & $T_2$ &Type$^{\,1}$&  $i$ [deg]   &  $\Omega_1$   & $\Omega_2$    & $L_1$ [\%] & $L_2$ [\%] & $L_3$ [\%] \\
  \hline\noalign{\smallskip}
 $\#$0850 &  32000  & 28222 (450) &    D       & 60.44 (0.76) & 4.160 (0.040) & 4.265 (0.039) & 50.4 (1.9) & 36.9 (0.7) & 12.7 (1.2) \\ %rms 114.8-
 $\#$1029 &   9700  & 10183 (95)  &   OC       & 78.19 (0.59) & 3.692 (0.019) &  --           & 47.3 (0.7) & 51.5 (0.8) & 1.2  (1.0) \\ %rms 5030-5004-
 $\#$1403 &  25000  & 24743 (651) &    D       & 62.58 (1.08) & 3.898 (0.057) & 4.236 (0.095) & 22.9 (1.3) & 17.4 (2.1) & 59.7 (3.2) \\ %rms 98.5-96.5-94.2-92.2-
 $\#$2036 &  15000  & 14406 (420) &    D       & 85.04 (0.85) & 5.341 (0.139) & 6.035 (0.210) & 16.9 (0.9) & 11.6 (0.7) & 71.5 (9.7) \\ %rms 9843-9831-9686-9655-9637-9625-
 $\#$2119 &  33800  & 31227 (368) &    D       & 67.14 (0.55) & 4.962 (0.082) & 3.964 (0.015) & 27.6 (0.7) & 53.8 (1.4) & 18.6 (2.6) \\%jiz analyzovany system: 2005MNRAS.357..304H, %rms 213-257-...
 $\#$2272 &  29000  & 24948 (337) &   SD       & 60.09 (0.68) & 3.943 (0.052) & --            & 38.0 (1.7) & 33.8 (1.1) & 28.2 (1.7) \\ %rms 182-
 $\#$2315 &  16000  & 13071 (245) &    D       & 76.24 (0.76) & 5.015 (0.098) & 5.245 (0.085) & 55.2 (1.1) & 36.4 (2.3) & 8.4  (1.9) \\ %rms 1414-1405-
 $\#$2639 &  26000  & 25920 (312) &    D       & 84.57 (0.89) & 4.146 (0.049) & 6.431 (0.088) & 30.3 (4.4) &  9.5 (2.0) & 60.2 (6.7) \\ %rms 1007-990-927-910-
 $\#$3910 &  41000  & 40218 (489) &    D       & 72.54 (0.35) & 4.759 (0.041) & 4.923 (0.024) & 50.9 (1.2) & 44.7 (0.9) & 4.4  (0.9) \\ %studie:  2012ApJ...748...96M :  sp O5+O7, triple lines!!!  %rms 295-291-
 $\#$3968 &  10000  &  9241 (101) &   SD       & 62.08 (0.19) & 3.713 (0.015) & --            & 53.3 (0.9) & 46.7 (1.2) & 0.0        \\ %rms 158-
 $\#$4013 &  10000  &  8839  (74) &   OC       & 73.33 (0.51) & 3.708 (0.030) & --            & 55.2 (0.7) & 44.8 (0.7) & 0.0        \\ %rms 340-
 $\#$4024 &  26000  & 25697 (345) &    D       & 85.15 (1.08) & 3.754 (0.049) & 3.921 (0.055) & 26.7 (2.1) & 22.6 (0.9) & 50.6 (1.7) \\ %rms 156-154-148-138-132-111-
 $\#$4662 &  33000  & 30078 (389) &    D       & 66.92 (1.01) & 4.438 (0.051) & 4.747 (0.061) & 35.1 (2.0) & 23.9 (2.3) & 41.0 (4.2) \\ %rms 133-126-
 $\#$4690 &  41000  & 39034 (392) &   OC       & 58.44 (0.49) & 3.129 (0.027) & --            & 50.5 (2.3) & 31.1 (0.8) & 18.4 (3.4) \\%jiz analyzovany system:  2003MNRAS.341..583M  %rms 114-107-101-
  \noalign{\smallskip}\hline
\end{tabular}
\begin{minipage}{0.9\textwidth}
% \scriptsize
Notes: [1] - D=Detached, OC=Overcontact, SD=Semidetached,
\end{minipage}
\end{table*}

\begin{table*}
\caption{The parameters of the third-body orbits for the individual systems.} \label{OCparam}
 \scriptsize
\begin{tabular}{ccccccccccc}
\hline\hline\noalign{\smallskip}
  System  & $HJD_0    $  &     $P$        &  $A$        &  $\omega$    & $P_3$      & $T_0$ [HJD] &      e      & $f(m_3)$   & $P_3^2/P$ \\   %pocet minim
          & (2450000+)   &     [days]     & [days]      &   [deg]      & [yr]       & (2400000+)  &             & $[M_\odot]$& [yr]      \\
 \noalign{\smallskip}\hline\noalign{\smallskip}
 $\#$0850 & 2088.833 (4) & 1.0028183 (12) & 0.0252 (28) &   4.0 (12.1) & 17.3 (3.2) & 84108 (925) & 0.465 (24)  & 0.398 (21) &  109223   \\
 $\#$1029 & 1176.969 (2) & 0.3766298 (3)  & 0.0111 (4)  & 156.5 (17.0) & 14.4 (0.6) & 54399 (204) & 0.264 (59)  & 0.038  (1) &  200598   \\
 $\#$1403 & 2121.714 (1) & 0.8685842 (5)  & 0.0092 (8)  & 256.6 (10.9) &  3.3 (0.1) & 52142 (28)  & 0.002 (3)   & 0.359  (2) &    4671   \\
 $\#$2036 & 1173.340 (2) & 1.2537097 (10) & 0.0128 (18) &   8.3 (15.1) &  4.6 (0.1) & 53170 (69)  & 0.735 (123) & 1.571 (514)&    6188   \\
 $\#$2119 & 2162.710 (2) & 2.1763852 (22) & 0.0162 (11) & 193.1 (23.2) &  5.3 (0.3) & 75476 (192) & 0.001 (140) & 0.800 (0.1)&    4670   \\
 $\#$2272 & 2123.872 (3) & 3.8209876 (46) & 0.0152 (41) & 112.7 (32.0) &  7.6 (0.4) & 74458 (136) & 0.004 (2)   & 0.319 (20) &    5487   \\
 $\#$2315 & 1173.514 (1) & 1.1268622 (3)  & 0.0038 (12) & 311.3 (16.5) &  9.7 (0.7) & 51806 (239) & 0.902 (191) & 0.006 (1)  &   30304   \\
 $\#$2639 & 1177.245 (2) & 1.1879433 (9)  & 0.0202 (11) &  65.2 (10.3) &  3.6 (0.1) & 52556 (30)  & 0.599 (90)  & 3.555 (52) &    4058   \\ %PREPOCIST????
 $\#$3910 & 2054.739 (4) & 2.3548423 (30) & 0.0115 (8)  &   0.5 (13.2) &  4.6 (0.1) & 66944 (27)  & 0.208 (15)  & 0.390 (17) &    3350   \\
 $\#$3968 & 2053.089 (1) & 0.2911953 (2)  & 0.0034 (2)  & 204.4 (19.7) &  2.0 (0.1) & 75323 (5)   & 0.002 (2)   & 0.051 (1)  &    5166   \\
 $\#$4013 & 3186.271 (1) & 0.2805165 (2)  & 0.0108 (4)  & 231.4 (13.9) & 17.1 (0.6) & 65028 (198) & 0.203 (58)  & 0.023 (2)  &  382833   \\[1mm]
 $\#$4024 & 2121.764 (3) & 1.1230775 (13) & 0.0087 (4)  & 216.8 (18.9) &  2.2 (0.1) & 50281 (58)  & 0.033 (10)  & 0.748 (5)  &    1511   \\
 $\#$4024 &              &                & 0.0306 (5)  &  23.4 (10.0) & 19.2 (0.4) & 43123 (354) & 0.283 (32)  & 0.453 (3)  &  119444   \\[1mm]
 $\#$4662 & 2141.146 (8) & 1.1670005 (39) & 0.0476 (19) &   0.0 (5.9)  & 16.3 (0.5) & 56058 (49)  & 0.566 (40)  & 3.755 (50) &   83160   \\
 $\#$4690 & 2130.790 (3) & 2.2061051 (32) & 0.0102 (18) &  62.8 (26.6) & 12.4 (1.7) & 82201 (627) & 0.249 (149) & 0.037 (2)  &   25529   \\ \hline
 \noalign{\smallskip}\hline
\end{tabular}
\end{table*}

\begin{figure}
  \centering
  \includegraphics[width=0.43\textwidth \vskip 1mm \hskip 2mm]{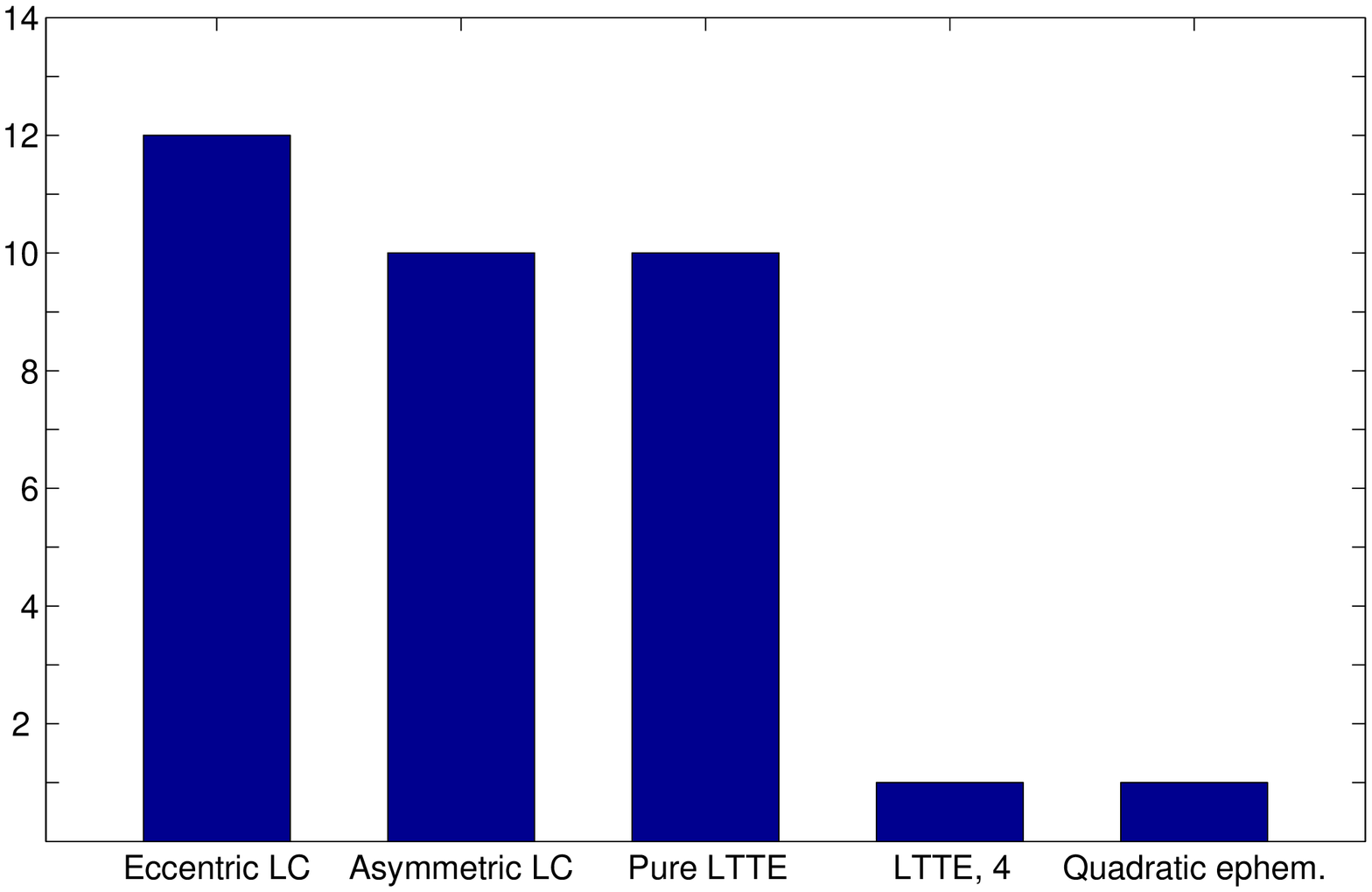}
  \caption{Number of system of each kind from our set of 100 brightest SMC eclipsing binaries (see the text for details).}
  \label{FigNumbers}
\end{figure}

\section{A little statistics} \label{stat}

We have manually checked the first one hundred brightest eclipsing binaries from the OGLE III
catalogue located in the SMC or its vicinity. These brightest stars have the largest probability to
contain the third components (see e.g. \citealt{2013ARA&A..51..269D}). Hence, we tested the
hypothesis of period changes and selected the most pronounced examples of periodic modulation of
the orbital periods. Moreover, for some other systems also the other phenomena were detected. These
findings are summarized below in Fig. \ref{FigNumbers}.

Among these one hundred stars we have found 11 systems showing LTTE modulation (see the previous
Section), twelve systems with obviously eccentric LC, and 10 systems with obviously asymmetric LC
shape. One system was classified as a mass-transfer system due to the rapid quadratic ephemerides
(OGLE-SMC-ECL-1232), however it is not sure whether such a behaviour is not only a very long-term
periodic variation. The rest of the stars show either no variation of their orbital periods or
their modulations were still questionable and more data are needed for a final confirmation.

% obviously eccentric LC : 12
% obviously asymmetric LC : 10 (vcetne 0277)
% LITE pure : 10
% LITE 4 body 1
% MT pure : 1
%
%grafika=[12,10,10,1,1];

\begin{table}
% \centering
% \tiny
% \scalebox{0.9}{
 \begin{minipage}{85mm}
  \caption{Heliocentric minima of the systems used for the analysis.} \label{MINIMA}
 %  \tiny
  \begin{tabular}{@{}l l l l l l@{}}
\hline
Star & HJD - 2400000 & Error & Type & Filter & Source\\
 \hline
 $\#$0850 & 50751.05396 & 0.00547 & Prim & I &  OGLE II \\
 $\#$0850 & 50751.55310 & 0.00331 & Sec  & I &  OGLE II \\
 $\#$0850 & 51098.02700 & 0.00086 & Prim & I &  OGLE II \\
 $\#$0850 & 51098.52062 & 0.00408 & Sec  & I &  OGLE II \\
 \dots \\
  \hline
 \end{tabular}
 \scriptsize This table is available in its entirety as a machine-readable table. A portion is shown here for
 guidance regarding its form and content.
\end{minipage}%}
%\begin{minipage}{0.9\textwidth}
\end{table}

\section{Results and discussion} \label{discussion}

The methods for analysing eclipsing binary are nowadays classical and used almost routinely.
However, any such kind of analysis can still bring new and surprising results, especially when
applying to some new group of targets. And this can be the case also for this study, which presents
the first analysis of the period changes of binaries located in the SMC galaxy.

The classical hypothesis of the light-travel time effect was applied to the 14 selected eclipsing
systems from the SMC and we found out that there are probably the third components with rather
short orbital periods of couple of years. Moreover, one system was also classified as a candidate
for a quadruple. Our focus on the bright (i.e. massive) stars was justified due to the strong
correlation between the multiplicity and the masses, as discovered via a study of galactical
populations (see e.g. \citealt{2010ApJS..190....1R}). However, at this point it is necessary to
emphasize that it is not clear whether all of these stars belong to the SMC galaxy or is it just a
coincidence that they lie in the same direction and are members of the Milky Way (however, it is
very unlikely because all of them have the distance moduli $>$10~mag).

The presented analysis showed that the predicted third bodies found via the period changes have
rather high masses in general, but this is due to the fact that also the eclipsing binary
components are massive ones. Another effect which can also play a role is a fact that only a
modulation with higher amplitudes are being detected in the data we were using from various
databases. We deal here with the stars of the highest luminosity, the highest mass and of the
earliest spectral type in the SMC, hence also the third bodies should be massive. This finding was
supported by the fact that also large fractions of the third light were usually detected in the
light curve solutions and these two numbers well coincide with each other for most of the systems
in our sample.

Obviously, we can ask whether in galaxies like SMC is e.g. the multiplicity fraction the same as
derived in our Milky Way. Certainly, there arises several selection effects of our method and also
some limitations of the available data. First of all, we have only very limited data coverage, of
about 20 years in maximum. Hence, also our estimated periods of the third bodies should be of
several years to two decades. Shorter orbital periods are rather hard to detect in the available
data. From different investigations (e.g. \citealt{1991A&A...248..485D},
\citealt{2008MNRAS.389..925T}, or \citealt{2010ApJS..190....1R}) there arises that the period
distribution of the outer orbits tend to have its maximum even at longer periods. Therefore, we are
able to detect here only a small fraction of the true triple systems (tip of the iceberg). Another
selection effect should arise from the fact that our method prefers more massive tertiary
components (causing larger LTTE amplitudes), while the less massive probably remain undetected.
Nevertheless, we can still summarize that our derived multiplicity fraction of about 10\% is very
similar to that one found by \cite{2016MNRAS.455.4136B} for the galactic population located in the
Kepler field. The mass distribution of the two samples is very different (hence also its
multiplicity), but with the super precise Kepler data one is able to detect also such triples,
which would easily be missed by this method in our sample.

Another finding resulted from our study is the fact that the gravity perturbing effects of the
third bodies on the eclipsing binary orbits are generally only very weak. Besides the classical
LTTE effect and its dynamical analogy for much closer systems \citep{2016MNRAS.455.4136B}, also the
longer-period effects should be present in the three-body system. This is for example the
precession of both orbits due to the third body. Most pronounced is typically the change of
inclination of the inner eclipsing pair. We are aware of the fact that such systems (called
transient eclipsing binaries) exist even in the OGLE database, see e.g. \cite{2011AcA....61..103G}
with 17 such systems located in LMC, or new study by \citep{Jurysek} with a few dozens of such
systems. However, the problem is that this effect depends on the period ratio $P_3^2/P$
\citep{1975A&A....42..229S}, see the last column in Table \ref{OCparam}, where one can see that
this effect is only very slow to hope finding any evidence for that in upcoming years in our
systems. Also the interferometry is out of the game, therefore only the spectroscopy can be used as
an independent method for finding the additional bodies in the selected systems.

\section{Conclusion}

A set of fourteen luminous eclipsing binaries in SMC were analysed resulting in a third-body
hypothesis. Some dedicated observation of these systems would be very welcome. This especially
applies for the interesting quadruple system OGLE-SMC-ECL-4024, or the systems with very short
period $p_3$ and high level of the third light (like $\#$1403, $\#$2036, or  $\#$2639).

\section*{Acknowledgements}
We do thank the {\sc MACHO} and {\sc OGLE} teams for making all of the observations easily public
available. This work was supported by the Czech Science Foundation grant no. GA15-02112S, and also
by the grant MSMT INGO II LG15010. We are also grateful to the ESO team at the La Silla Observatory
for their help in maintaining and operating the Danish telescope. The following internet-based
resources were used in research for this paper: the SIMBAD database and the VizieR service operated
at the CDS, Strasbourg, France, and the NASA's Astrophysics Data System Bibliographic Services.

%%%%%%%%%%%%%%%%%% APPENDICES %%%%%%%%%%%%%%%%%%%%%
%
%\appendix

% Don't change these lines
\bsp    % typesetting comment
\label{lastpage}

\begin{thebibliography}{99}
 \bibitem[\protect\citeauthoryear{Borkovits et al.}{2016}]{2016MNRAS.455.4136B} Borkovits, T., Hajdu, T., Sztakovics, J., et al.\ 2016, MNRAS, 455, 4136
 \bibitem[\protect\citeauthoryear{Davies et al.}{2015}]{2015ApJ...806...21D} Davies, B., Kudritzki, R.-P., Gazak, Z., et al.\ 2015, ApJ, 806, 21
 \bibitem[\protect\citeauthoryear{Duch{\^e}ne \& Kraus}{2013}]{2013ARA&A..51..269D} Duch{\^e}ne, G., \& Kraus, A.\ 2013, ARA\&A, 51, 269
 \bibitem[\protect\citeauthoryear{Duquennoy \& Mayor}{1991}]{1991A&A...248..485D} Duquennoy, A., \& Mayor, M.\ 1991, A\&A, 248, 485
 \bibitem[\protect\citeauthoryear{Faccioli et al.}{2007}]{2007AJ....134.1963F} Faccioli, L., Alcock, C., Cook, K., et al.\ 2007, AJ, 134, 1963
 \bibitem[\protect\citeauthoryear{Gies et al.}{2015}]{2015AJ....150..178G} Gies, D.~R., Matson, R.~A., Guo, Z., et al.\ 2015, AJ, 150, 178
 \bibitem[\protect\citeauthoryear{Graczyk et al.}{2011}]{2011AcA....61..103G} Graczyk, D., Soszy{\'n}ski, I., Poleski, R., et al.\ 2011, AcA, 61, 103
 \bibitem[\protect\citeauthoryear{Hilditch et al.}{2005}]{2005MNRAS.357..304H} Hilditch, R.~W., Howarth, I.~D., \& Harries, T.~J.\ 2005, MNRAS, 357, 304
 \bibitem[\protect\citeauthoryear{Hong et al.}{2016}]{2016MNRAS.460..650H} Hong, K., Lee, J.~W., Kim, S.-L., Koo, J.-R., \& Lee, C.-U.\ 2016, MNRAS, 460, 650
 \bibitem[\protect\citeauthoryear{Irwin}{1959}]{Irwin1959} {Irwin}, J.~B. 1959, AJ, 64, 149
 \bibitem[\protect\citeauthoryear{Jury\v{s}ek}{2017}]{Jurysek} Jury\v{s}ek, J., Zasche, P., Wolf, M., et al. 2017, A\&A, submitted
 \bibitem[\protect\citeauthoryear{Kallrath \& Milone}{2009}]{2009ebs..book.....K} Kallrath, J., \& Milone, E.~F.\ 2009, ''Eclipsing Binary Stars: Modeling and Analysis: Astronomy and Astrophysics Library'', Springer-Verlag New York
 \bibitem[\protect\citeauthoryear{Massey}{2002}]{2002ApJS..141...81M} Massey, P.\ 2002, ApJS, 141, 81
 \bibitem[\protect\citeauthoryear{Massey et al.}{2012}]{2012ApJ...748...96M} Massey, P., Morrell, N.~I., Neugent, K.~F., et al.\ 2012, ApJ, 748, 96
 \bibitem[\protect\citeauthoryear{Mayer}{1990}]{Mayer1990} Mayer, P.\ 1990, BAICz, 41, 231
 \bibitem[\protect\citeauthoryear{Pawlak et al.}{2013}]{2013AcA....63..323P} Pawlak, M., Graczyk, D., Soszy{\'n}ski, I., et al.\ 2013, AcA, 63, 323
 \bibitem[\protect\citeauthoryear{Pawlak et al.}{2016}]{2016AcA....66..421P} Pawlak, M., Soszy{\'n}ski, I., Udalski, A., et al.\ 2016, AcA, 66, 421
 \bibitem[\protect\citeauthoryear{Pr{\v s}a \& Zwitter}{2005}]{2005ApJ...628..426P} Pr{\v s}a, A., \& Zwitter, T.\ 2005, ApJ, 628, 426
 \bibitem[\protect\citeauthoryear{Raghavan et al.}{2010}]{2010ApJS..190....1R} Raghavan, D., McAlister, H.~A., Henry, T.~J., et al.\ 2010, ApJS, 190, 1
 \bibitem[\protect\citeauthoryear{Southworth}{2012}]{2012ocpd.conf...51S} Southworth, J.\ 2012, Orbital Couples: Pas de Deux in the Solar System and the Milky Way, 51
 \bibitem[\protect\citeauthoryear{S\"oderhjelm}{1975}]{1975A&A....42..229S} S\"oderhjelm, S.\ 1975, A\&A, 42, 229
 \bibitem[\protect\citeauthoryear{Terrell \& Wilson}{2005}]{2005Ap&SS.296..221T} Terrell, D., \& Wilson, R.~E.\ 2005, Ap\&SS, 296, 221
 \bibitem[\protect\citeauthoryear{Tokovinin}{2008}]{2008MNRAS.389..925T} Tokovinin, A.\ 2008, MNRAS, 389, 925
 \bibitem[\protect\citeauthoryear{Udalski et al.}{1998}]{1998AcA....48..147U} Udalski, A., Szymanski, M., Kubiak, M., et al.\ 1998, AcA, 48, 147
 \bibitem[\protect\citeauthoryear{van Hamme}{1993}]{1993AJ....106.2096V} van Hamme, W.\ 1993, AJ, 106, 2096
 \bibitem[\protect\citeauthoryear{Westerlund}{1997}]{1997macl.book.....W} Westerlund, B.~E.\ 1997, ''The Magellanic Clouds'', Cambridge University Press, UK
 \bibitem[\protect\citeauthoryear{Wilson \& Devinney}{1971}]{1971ApJ...166..605W} Wilson, R.~E., \& Devinney, E.~J.\ 1971, ApJ, 166, 605
 \bibitem[\protect\citeauthoryear{Wyrzykowski et al.}{2004}]{2004AcA....54....1W} Wyrzykowski, L.,Udalski, A., Kubiak, M., et al.\ 2004, AcA, 54, 1
 \bibitem[\protect\citeauthoryear{Zaritsky et al.}{2002}]{2002AJ....123..855Z} Zaritsky, D., Harris, J., Thompson, I.~B., Grebel, E.~K., \& Massey, P.\ 2002, AJ, 123, 855
 \bibitem[\protect\citeauthoryear{Zasche}{2011}]{2011IBVS.5991....1Z} Zasche, P.\ 2011, Information Bulletin on Variable Stars, 5991, 1
 \bibitem[\protect\citeauthoryear{Zasche et al.}{2014}]{2014A&A...572A..71Z} Zasche, P., Wolf, M., Vra{\v s}til, J., et al.\ 2014, A\&A, 572, A71
 \bibitem[\protect\citeauthoryear{Zasche et al.}{2016}]{2016A&A...590A..85Z} Zasche, P., Wolf, M., Vra{\v s}til, J., Pilar{\v c}{\'{\i}}k, L., \& Jury{\v s}ek, J.\ 2016, A\&A, 590, A85
\end{thebibliography}
\end{document}